\documentclass[aps,prb,groupedaddress,twocolumn,10pt]
{revtex4-1}

\usepackage{graphicx}

\begin{document}

\newcommand{\be}{\begin{equation}}
\newcommand{\ee}{\end{equation}}
\newcommand{\bea}{\begin{eqnarray}}
\newcommand{\eea}{\end{eqnarray}}

\title{Simulation of holographic correspondence in flexible graphene}

\author{D. V. Khveshchenko}

\affiliation{Department of Physics and Astronomy, University of North Carolina, Chapel Hill, NC 27599}

\begin{abstract}
In the spirit of the generalized holographic conjecture, we explore a relationship between the bulk and boundary properties of non-interacting massive Dirac fermions living on a flexible surface, such as a sheet of graphene. We demonstrate that the boundary correlations can mimic those normally found in the system of one-dimensional interacting fermions, a specific form of such phantom interaction being determined by the bulk geometry. This geometrical interpretation of the boundary interaction effects offers a new insight into the possible origin of the more sophisticated types of holographic correspondence and suggests potential ways of visualizing 'analogue holography' in the experimentally viable environments. 
\end{abstract}

\maketitle

\nopagebreak
In the past few years, there has been a tremendous activity in the area of holographic correspondence. This notion was first put forward with regard to the conjectured (possibly, exact) relationship between certain highly symmetric relativistic $4d$ gauge field and $5d$ string theories, in which context it is commonly referred to as '$AdS/CFT$ duality' \cite{AdS}. 

More recently, the holographic ideas were further extended 
by abandoning nearly all of the original stringent symmetry conditions  
in the hope of still capturing some key aspects of the underlying correspondence. 
Specifically, it was conjectured to remain applicable to a broad variety of non-symmetric and/or non-relativistic systems  which, incidentally, would be of interest to condensed matter physics. 

Such a drastic ('non-$AdS$/non-$CFT$') generalization 
was motivated, among other things, by a growing realization that the limited set of the classic '$AdS$ black brane' geometries utilized in the early work \cite{NFL} appears to be much too restrictive, thus allowing access to, essentially, just one specific type of all the possible non-Fermi liquid (NFL) 
compressible states of fermions in $d$ spatial dimensions. 

Namely, the historic Reissner-Nordstrom (RN) solutions to the coupled 
Einstein-Maxwell equations which asymptotically approach the 
$AdS_{d+2}$ and $AdS_2\times R^d$ geometries in the ultraviolet (UV) and infrared (IR) limits, respectively, was shown to invariably result in the behavior dubbed as 'semi-local criticality' \cite{NFL}. This particular regime is characterized by the fermion propagator $G(\omega, k)$ demonstrating a non-trivial frequency, yet mundane momentum, dependence, 
as manifested by the self-energy $\Sigma(\omega,k)\sim\omega^{\nu_k}$ where $\nu_k$ is a regular function with no singularities at the putative Fermi surface(s). 

Superficially, this 'generalized marginal Fermi liquid' bears a certain resemblance to that found in some of the heavy fermion compounds. 
However, it turns out to be plagued with such starkly spurious features 
as multiple Fermi surfaces or non-vanishing zero 
temperature entropy and, therefore, could only be identified 
with some intermediate, rather than the true asymptotic IR, regime. 

Regardless of the physical relevance of the above scenario, though, it would be interesting to find potential gravity duals for other types of, both, the documented as well as suspected NFL states of correlated fermions. Of a particular interest are those 'strange' Fermi (non-relativistic) and Dirac (pseudo-relativistic) metals where both, the frequency and momentum, dependence of the fermion propagator would be markedly non-trivial. 

The recent efforts in that direction produced a number of prospective geometries, including the Schroedinger, Lifshitz, and, especially, hyperscaling violating ones. Such metrics were found among the 'electron star' solutions of the minimal Einstein-Maxwell theory with back-reaction of the fermionic matter included, 
as well as those of the alternate dilaton, massive vector field, and Horava gravity theories \cite{hyperscaling}. 

However, while the use of such geometries can greatly expand the list of potentially attainable boundary NFL theories, it still does not clarify the status of the generalized holographic conjecture itself. 

In this paper, we add to the ongoing discourse by demonstrating that a certain form of the bulk-to-boundary correspondence might indeed be common for the systems in question, regardless of the presence of any extended (super)symmetries or a lack thereof. 

To that end, we consider a number of examples where the physical edges of the system of non-interacting $2d$ Dirac fermions propagating in curved spaces exhibit some non-trivial properties that would be typically attributed to the effects of certain $1d$ interactions. In contrast to the aforementioned 'semi-local' scenario \cite{NFL}, though, we show that in all these cases it is the momentum, rather than the frequency, dependence of the propagator $G(\omega,k)$ of the 
$1d$ boundary fermions that gets altered.
 
It might be tempting to view this form of correspondence as yet another aspect of the general Einstein's equivalence principle, according to which the effect of a curved metric can be described as a certain interaction. Also, this observation suggests that, barring all the practical challenges, it might be possible to observe the signatures of such a relationship in custom-deformed flakes of graphene grown on commensurate substrates ($h-BN$ or alike), the latter endowing the bulk fermions with a finite mass via hybridization \cite{hBN}. 

The generally covariant action describing 
the kinematics of massive $d+1$-dimensional Dirac-like electronic 
excitations at zero temperature and density 
propagating in a curved geometrical background 
reads (hereafter the Fermi velocity is chosen to be unity)
\be
S=\int drdtd^dx{\sqrt {|det {\hat g}|}}{\bar \psi}\gamma^ae^\mu_a(i\partial_\mu +
{i\over 8}\omega_\mu^{bc}[\gamma_b,\gamma_c]+A_\mu-m)\psi
\ee 
In the case of a flexible membrane ($d=1$), such as a strained sheet of graphene,   
the vielbein $e^\mu_a$ determining the induced metric 
$g_{\mu\nu}=e_a^\mu e^b_\nu\eta_{ab}$, vector potential $A_\mu$, and spin connection $\omega^{ab}_{\mu}$ can be expressed in terms of the local lattice displacement and its derivatives \cite{vozmediano}. 

In what follows, we consider a class of static 
rotationally-invariant diagonal metrics represented by the interval
\be
ds^2=-f(r)dt^2+g(r){dr^2}+q(r){d {\vec x}^2}
\ee
where $r\leq R$ is the ('holographic') radial coordinate, and perform the Wick rotation of the time variable (measured in the laboratory frame), $t\to i\tau$, thereby switching to the Euclidean signature.

In the holographic prescription \cite{AdS}, 
the (retarded) boundary propagator $G(\omega,{k})$, where $k=|{\vec k}|$, 
of a spin-$s$ probe particle subject to the bulk metric (2) would be obtained by 
finding a zero-energy solution of the radial 
wave equation with some effective potential $V(r,\omega, k)$  (see Eq.(7) below)
\be
{\partial^2 \psi(r,\omega,{k})\over {\partial r}^2}=V(r,\omega,{k})\psi(r,\omega,{k}) 
\ee
that satisfies the in-falling boundary condition at the IR cutoff $r=a$ \cite{NFL}. Expanding the thus-obtained solution in the opposite (UV) regime, i.e. near the boundary at $r=R\gg a$, over the functions $\psi_{\pm}(r,\omega,k)$ which are chosen as normalizable and un-normalizable, respectively, one then reads off the boundary propagator as the reflection coefficient for the incident radial wave
\be
G(\omega,k)={\psi_+(r,\omega,k)\over \psi_-(r,\omega,k)}|_{r\to R}
\ee
Despite its rather specialized construction (that is neither unique, nor non-debatable \cite{AdS,NFL}), the propagator (4) shares its singular dependence on $\omega$ and $k$ (if any) with the more conventional Green function of the boundary problem for the Sturm-Liouville equation (3) 
\be
G(r,r^\prime,\omega,k)={\theta(r-r^\prime)
\psi_+(r)\psi_-(r^\prime)+(r\leftrightarrow r^\prime)\over
\psi_-{d\psi_+\over dr}-\psi_+{d\psi_-\over dr}}
\ee
when the arguments $r$ and $r^\prime$ simultaneously approach the boundary, 
$r, r^\prime\to R$. Notably, Eq.(5) represents the conventional 
bulk propagator of massive $2d$ Dirac fermions in 
a curved geometry without any specific reference to the holographic conjecture whatsoever. 

Although Eq.(3) can not be solved for generic gravitational 
backgrounds, in the regime $m(\tau, x)\gg 1$
one can resort to the semiclassical approach \cite{WKB} and choose
\be 
\psi_{\pm}(r,\omega,k)\sim {1\over V^{1/4}(r,\omega,k)}
e^{\mp\int^{R}_{r}dr^\prime{\sqrt {V(r^\prime,\omega,k)}}}
\ee
where  $r_t$ is the turning point defined as $V(r_t)=0$.

Moreover, to the leading order in $mR\gg 1$, the semiclassical effective potential in (3) appears to be independent of the probe's spin 
\be
V(r,\omega,k)=g(r)(m^2+{k^2\over q(r)}+{\omega^2\over f(r)})+\dots,
\ee  
where the dots stand for the subdominant $s$-dependent terms \cite{WKB}.

Thus, in the space-time domain the asymptotic large-scale decay 
\be 
G(\tau,x)\sim\exp(-S_0(\tau,x))
\ee
of the boundary propagators (4) and (5) for a field of any spin 
is governed by the classical action
\be
S(\tau, x)=m\int dr {\sqrt {g(r)+f(r)({d\tau/dr})^2+q(r)({d{\vec x}/dr})^2}} 
\ee
computed along the extremal path between the points $(0,0,R)$ and $(\tau,x,R)$. 

After being evaluated upon such a geodesic trajectory, Eq.(9) yields 
\be
S_0(\tau,x)=2m^2\int^{R}_{r_t} dr{{\sqrt g(r)}\over M(r)}
\ee
where $M(r)={\sqrt {m^2-{k^2/q(r)}-{\omega^2/f(r)}}}$ and
the factor of two accounts for 
the particle's radial excursion from $R$ down to $r_t$ and back.

In Eq.(10) the values of $\tau$ and $x$ are functions of the conserved canonical momenta $\omega$ and $k$ which must be determined from the equations of motion
\be
x=k\int^{R}_{r_t}{dr{\sqrt g(r)}\over q(r)M(r)},~~~
\tau=\omega\int^{R}_{r_t}{dr{\sqrt g(r)}\over f(r)M(r)}
\ee
As the simplest $2d$ geometry, we first consider a flat circle 
of radius $R$ with the line element 
\be
dl^2_{flat}=dr^2+r^2d\phi^2
\ee
its natural embedding into the $3d$ Euclidean space-time being given by the interval 
$ds^2=d\tau^2+dl^2$.
 
Switching from the in-plane angular variable $\phi$ to a (compactified) 
boundary coordinate $x=R\phi$ we obtain the momenta $k=m\cos(x/2R)$ and $\omega=m\tau/{\sqrt {\tau^2+4R^2\sin^2(x/2R)}}$. For the minimal path (chord) connecting two points on the circular boundary of radius $R$ Eq.(10) then yields 
\be
S_{flat}(\tau,x)=m{\sqrt {\tau^2+4R^2\sin^2(x/2R)}}
\ee
which is characteristic of a non-interacting massive field. The corresponding  
dynamical critical exponent is $z=1$, as indicated by the 
relative scaling between the spatial and temporal coordinates, $\tau\sim x^z$.

Next, we consider a surface of revolution (SOR) described by the line element
\be
dl^2_{sor}=dr^2[1+({\partial h(r)\over \partial r})^2]+r^2d\phi^2
\ee
where $h(r)$ is the vertical displacement out of the $x-y$ plane.

For a graphene sheet shaped as a funnel, 
$h(r)\sim (R/r)^\eta$ for $r\geq a$, the 'warp factor' $g(r)\sim 1/r^{2\eta+2}$
diverges at small $r$. One then obtains 
\be
S_{sor}(\tau,x)=m{\sqrt {\tau^2+(Rx^{\eta})^{2/(\eta+1)}}}
\ee
which reveals an unconventional behavior of the edge propagator (8) as a function of the distance along the edge. In contrast, its temporal 
dependence remains trivial, thus implying the 'holographic' value of the
dynamical critical exponent $z_{hol}={\eta/(\eta+1)}$.

It is instructive to compare the asymptotic (15) with the propagator of $1d$ fermions interacting via a pairwise potential $U(x)\sim 1/x^\sigma$ with $\sigma<1$. The latter can be evaluated with the use of the standard bozonization technique \cite{boson}. To leading approximation, the chiral (left/right moving) components of that propagator read
\be
G^{\pm}_{bos}(\tau,x)\sim\exp[-\int {dk\over 2\pi}{2+U_k\over \epsilon_k}(1-e^{\pm ikx-\epsilon_k t})]
\ee
where $\epsilon_k=k{\sqrt {1+U_k}}$ is the $1d$ plasmon dispersion, suggesting the dynamical exponent $z_{bos}=(1+\sigma)/2$. 

Matching the large-$x$ asymptotics, one finds that Eq.(15)
mimics the spatial decay of the propagator (16), provided that
 $\eta=(1-\sigma)/(1+\sigma)$. 
However, comparing the long-$\tau$ asymptotics we find them to be incompatible, 
as the former suggests $z_{hol}=(1-\sigma)/2$, in contrast with the above $z_{bos}$ for all $\sigma\neq 0$. 

In fact, the long-time behavior in the boundary theory would not be readily 
recoverable with the use of any bulk metric with a constant $f(r)$ (we expound on this point below). By contrast, in the 'semi-local' $AdS_2$ regime \cite{NFL,WKB} the counterpart of Eq.(15), $S_{s-l}(\tau, x)={\sqrt {(1-\nu_0)^2(\ln\tau/a)^2+m^2x^2}}$, manifests a predominantly temporal character of the NFL correlations in that case. 

Another instructive example is provided by the line element  
\be
dl^2_{log}=dr^2+R^2\exp({-2(r/R)^\lambda})d\phi^2
\ee
For $\lambda=1$ Eq.(17) represents a $2d$ surface of constant negative (Gaussian) curvature known as 'Beltrami trumpet' which,  
using the parametrization $\rho=R\ln(R/r)$, can be transformed into the 'Lobachevsky plane', $dl^2={d\rho^2/\rho^2+\rho^2d\phi^2}$.
Notably, its conformally flat embedding into the physical $3d$ space-time was also invoked in the recent discussions of the possibility of observing the analogue Unruh-Hawking effect in graphene \cite{iorio}. 

By computing (10) one obtains
\be
S_{log}(\tau,x)=m{\sqrt {\tau^2+R^2(\ln x/a)^{2/\lambda}}}
\ee
For $\lambda=1$ and at large $x$ the propagator (8) then decays algebraically, $G(0,x)\sim 1/x^{mR}$ which is reminiscent of the behavior found in the $1d$ Luttinger liquids \cite{boson}.

In contrast, for $\lambda\neq 1$ Eq.(18) yields a variety of stretched/compressed exponential  asymptotics which decay faster (for $\lambda<1$) or slower (for $\lambda>1$)
than any power-law. For instance, by choosing $\lambda=2/3$ one can simulate a faster-than-algebraic spatial decay, $G(0,x)\sim \exp(-const\ln^{3/2}x)$, in the $1d$ Coulomb gas ($\sigma=1$) which is indicative of the incipient formation of a $1d$ charge density wave \cite{boson}.

For other values of $\lambda$ Eq.(18) reproduces the behavior in the boundary theory governed by the interaction $U(x)\sim (\ln x)^{(2/\lambda)-3}/x$. Although the physical origin of such a bare potential would not be immediately clear, multiplicative logarithmic factors do routinely emerge in those effective $1d$ couplings that are associated with various marginally (ir)relevant two-particle 
('double-trace', in the $AdS/CFT$ context) operators \cite{boson}.

Turning now to the metrics with $f(r)\neq const$, one classic example is provided by the so-called BTZ solution \cite{BTZ}. 
In the UV limit, the non-rotating BTZ metric approaches the $AdS_3$ one  
\be
ds^2_{AdS}=(d\tau^2+dx^2)r^2+{dr^2\over r^2}
\ee
and, correspondingly, Eq.(10) produces the expressly Lorentz-invariant result
\be
S_{AdS}(\tau,x)=2mR\ln({{\sqrt {\tau^2+x^{2}}}\over a}+
{\sqrt {{\tau^2+x^{2}\over a^2}+1}})
\ee
which is in full agreement with the exact zero temperature boundary propagator (in the general 
case of a rotating BTZ solution the two chiral sectors have different temperatures \cite{BTZ}) 
\be
G_{AdS}(\tau,x)\sim {1\over (x-i\tau)^{2\Delta_+}(x+i\tau)^{2\Delta_-}}
\ee 
Notably, the exact left/right dimensions 
$\Delta_{\pm}=mR/2+1/2\pm 1/4$ of the boundary fermion operator satisfy the condition $\Delta_{+}+\Delta_{-}>1$. Therefore, the corresponding boundary theory can not be obtained from any short-ranged repulsive interaction $U(x)\sim\delta(x)$, in which case the corresponding Luttinger parameter would be restricted to the interval $1/2\leq K\leq 1$ \cite{boson}, thereby imposing the upper/lower bounds on the total conformal dimension,  
$1/2\leq \Delta_{+}+\Delta_{-}={1\over 4}(K+1/K)\leq 5/8$, contrary to the above.

Nevertheless, a power-law interaction potential 
$U(x)\sim 1/x^\sigma$ with $\sigma<1$ makes the Luttinger 
parameter $K$ momentum-dependent and can drive it all the way down to zero
for $k\to 0$, thereby raising the above upper bound. 
However, as already shown, this interaction results in a 
different, non-algebraic, $x$-dependence. 
Therefore, the boundary conformal field theory dual to the BTZ solution does not appear to have a microscopic realization in terms of any of the aforementioned pairwise potentials. 

In general, while a metric with $f(r)\neq const$ may not be 
readily attainable in the lab, it might still be possible to practically construct  
its conformal equivalent known as the Zermelo optical metric. 
It is worth mentioning, though, that the propagator of massive Dirac fermions would not remain invariant under the conformal transformation relating 
the two metrics. 
 
Specifically, under the parametrization $r=a\coth\rho/R$ the conformally flat BTZ solution reads 
\be
ds^2_{BTZ}={1\over \sinh^2\rho/R}(({a\over R})^2d\tau^2+{d\rho^2}
+a^2\cosh^2({\rho\over R}) d\phi^2)
\ee
and the corresponding optical interval is given by 
the expression in the brackets, its spatial part being readily identifiable 
as a hyperbolic pseudosphere whose practical realization
has also been envisioned in graphene \cite{iorio}. 

The above analysis can be extended in a number of ways. 
For one, the observed correspondence can be extended to the other physical observables of geometrical origin, one relevant example being entanglement entropy which is naturally 
associated with the area of a minimal surface \cite{AdS}.

Also, a non-trivial background metric can be complemented 
by various patterns of the vector potential $A_\mu$ which for
rotationally-invariant configurations amounts to substituting the frequency and momentum in Eq.(7) with $\omega-A_t(r)$ and $k-A_\phi(r)$. In the case of a graphene sheet,   
this (pseudo) electromagnetic potential represents both, elastic strain as well as extrinsic curvature associated with pentagonal/heptagonal and other localized structural defects \cite{vozmediano}.

Taken at their face value, our results demonstrate that certain seemingly interaction-like features can indeed emerge at the boundary of even a non-interacting bulk theory, provided that the latter is defined in a curved space. 
This observation suggests that some limited form of the holography-type relation might, in fact, be quite robust and hold regardless of whether or not the system in question is highly symmetrical, as per the original $AdS/CFT$ conjecture.

It remains to be seen, though, as to whether or not such an intrinsic form of the bulk-to-boundary relationship could account for any of the 
circumstantial evidence that was argued to support the 
generalized holographic conjecture \cite{AdS,NFL,hyperscaling,WKB}. 
However, a direct comparison is complicated by the fact that
most of the earlier results (with the exception of those precious few that pertain to the classic RN solutions at zero fermion density and/or mass and can be found explicitly in terms of the Heun functions \cite{NFL}) were obtained numerically and, therefore, do not 
offer a clear physical insight.
 
On the experimental side, the availability of (both, already existing and still emerging) flexible graphene devices and stress engineering techniques may allow for an experimental study of the analogue gravity duals of simulated $1d$ fermion systems. 
To that end, a proper choice of the substrate is instrumental for not only opening a bulk gap, but also for suppressing various edge magnetization effects which can further complicate matters. Conceivably, the boundary correlations can be probed with 
such relevant experimental techniques as 
time-of-flight, edge tunneling, and local capacitance measurements. 

Lastly, some form of the bulk-boundary duality
could also be anticipated in the properties of planar Dirac fermions residing on the surfaces of gapped $3d$ topological insulators subjected to stress.  
However, apart from the obvious complexity of engineering curved $3d$ spaces,
the $2d$ boundary systems are also known to be much 
less likely to exhibit any effects of the 
interactions, whether of a real or phantom (holographic) nature,
thus making their experimental detection significantly more intricate.

To summarize, we believe that establishing the true status of the generalized holographic conjecture would be far more important than continuing to apply it to the ever increasing 
number of model geometries and their physically obscure boundary duals.
In that regard, exploring even the simplest examples, such as that 
of the $3d$-$2d$ duality discussed in this work, should help one to clarify the minimal conditions required for the sought-after holographic relationship to hold, 
as well as the general physical reason(s) as to $why$ such a relationship 
could be expected in the first place. Besides, we argue that the already available 'massive Dirac metals' might offer a viable experimental playground 
for simulating various aspects of the holographic principle, thereby making it possible to test this tantalizing idea in the lab.  


\end{document}